# Dispersion from Diffuse Reflectors and its Effect on Terahertz Wireless Communication Performance

Russ Messenger, Karl Strecker, Sabit Ekin, *Member, IEEE* and John F. O'Hara, *Senior Member, IEEE*

*Abstract*—This work investigates the temporal dispersion of a wireless terahertz communication signal caused by reflection from a rough (diffuse) surface, and its subsequent impact on symbol error rate versus data rate. Broadband measurements of diffuse reflectors using terahertz time-domain spectroscopy were used to establish and validate a scattering model that uses stochastic methods to describe the effects of surface roughness on the phase and amplitude of a reflected terahertz signal, expressed as a communication channel transfer function. The modeled channel was used to simulate a quadrature phase shift keying (QPSK)-modulated wireless communication link to determine the relationships between symbol error rate and data rate as a function of surface roughness. The simulations reveal that surface roughness from wall texturing results in group delay dispersion that limits achievable data rate with low errors. A distinct dispersion limit in surface roughness is discovered beyond which unacceptable numbers of symbol errors begin to accrue for a given data rate.

## I. INTRODUCTION

THE continually growing need for available signal bandwidth to support higher data rates has been driving terahertz wireless communication research and technology for several years [1,2]. Data rates exceeding 10 Gbps may soon be required to support applications such as streaming high-definition video [3], telemedicine [4], online education [5] and machine-to-machine communication in the Internet of Things (IoT) [6], meaning even 5G (5th Generation) wireless systems cannot offer sufficient capacity.

The transition to terahertz frequencies introduces many new difficulties that could limit wireless communication system performance. Several of these derive from the small wavelength of terahertz radiation (0.03-3 mm for 10-0.1 THz). While this small wavelength offers the advantage of producing highly directive beams with small antennas, it also complicates providing adequate signal coverage [7,8] and managing multipath effects (e.g. fading [9]). These factors have been expected to initially limit terahertz wireless links to short range or line-of-sight (LOS) applications [10]. However, LOS communications may not always be possible or even desirable in many situations [11]. Therefore, a strong and growing incentive exists to investigate terahertz scattering from walls or other surfaces to reach intended recipients, thus taking advantage of existing obstacles to create non-line-of-sight (NLOS) communication channels.

Researched methods to overcome signal coverage limitations include the use of diffuse [11,12] and engineered [13,14] reflecting surfaces with the goal of *spatially* dispersing an otherwise highly directive beam to NLOS receivers. Investigated surfaces include common building materials such as plaster, wood, cinderblock, and glass [15,16] as well as engineered metasurfaces [14,17,18]. Many of these surfaces would be considered smooth, specular reflectors at microwave frequencies but have appreciable surface roughness at terahertz frequencies. The diffuse scattering of terahertz waves necessitates continued experimental characterization as well as new understanding and modifications to communication channel models.

Experimental and simulation studies have been undertaken in this area [8,19], leading to important improvements in terahertz scattering models. Sheikh *et al*. discusses two different models, the Beckmann-Kirchhoff and Effective Roughness models, showing that both are capable of predicting total received power of diffusely scattered terahertz waves [8]. Piesiewicz *et al*. developed modified Fresnel equations and ray tracing to simulate wall and ceiling scattering of terahertz radiation, quantifying absolute power and propagation patterns in indoor scenarios [19]. Priebe *et al*. measured the channel impulse response and transfer function of an indoor propagation environment, capturing power delay profile and reflection loss for wood, plaster, and plastic surfaces [7]. Dikmelik *et al*. introduced an analytical model to approximate reflections in the specular direction off rough surfaces leading to an explanation of fading, specifically the roll-off of spectral magnitude with increasing scattering surface height distribution [20]. Ma *et al*. and references therein comprise a review of several scattering measurements to date and Ma *et al*. also shows actual terahertz communication performance in NLOS applications using an amplitude shift keyed system at 100, 200, 300, and 400 GHz [16]. The results proved that low bit error rate is achievable at 1 Gbps over practically relevant angular distributions. While many of these studies mention multipath and delay spreading due to diffusing or rough surfaces, none has addressed the quantitative impact of *temporal* dispersion on the relationship between achievable data rate and symbol error rate (SER).

In this article, we use a combination of broadband, terahertz time-domain measurements and simulation to investigate the impact of group delay dispersion (GDD), induced by rough surface reflections, on symbol error rates at various high data rates. Since indoor walls and wall coverings are often textured for aesthetic appeal, there is practical value in exploring the effects of rough wall textures on the terahertz channel. Therefore, we employ textured drywall samples to act as scattering surfaces. Using a channel model based on reflection measurements from these textured samples, we simulate digital wireless communication performance with quadrature phase shift keying (QPSK) modulation at rates of 10-50 Gbd (gigabaud) using a carrier frequency of 0.25 THz. We find a distinct "dispersion limit" appears in the scattering surface height distribution beyond which GDD causes massive expansion of the SER, such that communication may become impractical at high data rates.

The rest of this manuscript is organized as follows. Section II discusses how the samples used for measurements were prepared and the methodology by which measurements were taken. Section III analyses the measurement results and describes how a channel model for simulation was developed based on the measured data. Section IV predicts possible terahertz wireless communication system limitations based of the simulation model. Finally, section V presents the conclusions and implications of this work.

## II. Measurements

### A. Sample Preparation

All of the drywall samples used in the experiments were cut from a single 0.5 inch thick panel of Georgia-Pacific ToughRock Drywall. The first sample consisted of an untextured 100 mm x 100 mm section of drywall, painted with an indoor, 100% acrylic paint manufactured by the Sherwin-Williams Company (Fig 1a). Additional drywall samples were cut to the same size and then coated with mixtures of the same paint, but containing various concentrations of aluminum (Al) powder. These metal "doped" paint samples were made using two different Al powders. The first was a fine, 3.0-4.5 micron spherical aluminum powder and the second was a coarse, 149-297 micron aluminum powder. Aluminum powder was selected to provide texture (for enhancing spatial dispersion) and to simultaneously increase the reflectivity of the surface, which would increase the received signal strength for NLOS communications. The paint was mixed with each Al powder in the following mass (powder) to volume (paint) ratios: 500 mg/mL, 250 mg/mL, 100 mg/mL, 50 mg/mL, 25 mg/mL, 10 mg/mL, and 5 mg/mL. The doped paints were applied with a paintbrush to the drywall samples in the same manner as the undoped paint. Given its larger particle size, the coarse Al powder paint produced a highly visible texture following application, as shown in Fig. 1b.

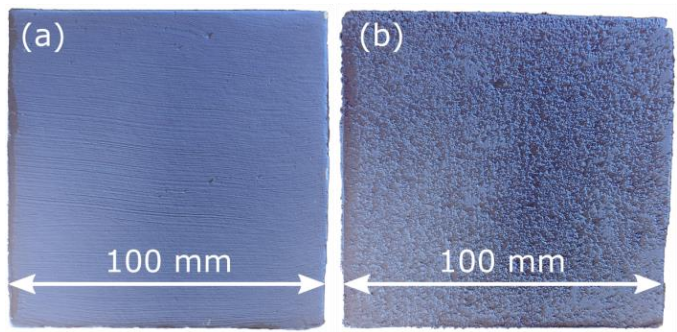

**Fig. 1:** Drywall samples with (a) undoped, and (b) 500mg/mL coarse Al powder doped paint.

### B. Measurement Methods

Reflection measurements were performed with terahertz time-domain spectroscopy (THz-TDS) [21], as illustrated in Fig. 2. The receiver is fiber-coupled to the gating laser, so it is capable of rotating around the *z*-axis without altering laser alignment or timing. This enables the measurement of the spatial dispersion of the scattered energy, however all measurements in this work utilized a specular arrangement (at 45° incidence angle) to maximize received signal power. The spatial dispersion was measured, but surprisingly, it was found to be modest relative to the normal expansion of the beam diameter due to free-space propagation.

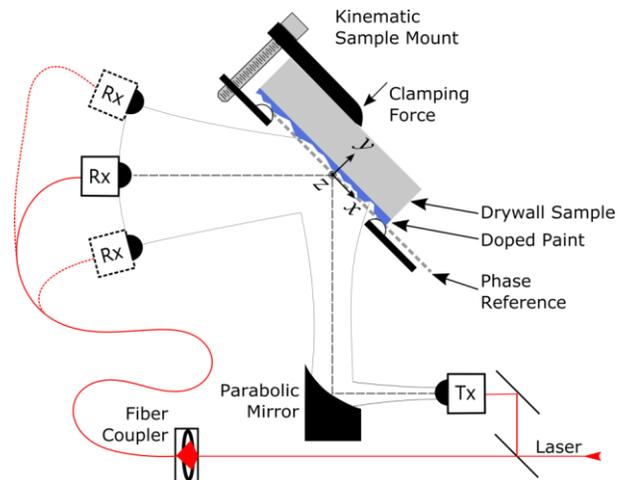

**Fig. 2:** THz-TDS system. Samples are mounted with the scattering face pressed lightly against kinematic stops to provide a constant phase reference, inasmuch as possible. The terahertz receiver is fiber-coupled and capable of rotating around the sample mount (*z*-axis) to measure scattering in the horizontal (*xy*) plane.

Reflection measurements were collected from both the doped and undoped drywall samples as well as a first-surface flat mirror, the last serving as a phase and amplitude reference for both samples. Drywall slabs with no paint and undoped paint served as control samples to isolate the effect of the paint alone. The reference mirror was measured between each sample to ensure minimum drift of the THz-TDS system during measurements. All samples and the reference mirror were mounted such that their front surfaces coincided with a single phase reference plane established by kinematic stops, as illustrated in Fig. 2. This was an imperfect scheme since the surface roughness of the samples compromised the performance of the kinematic stops. Nevertheless, the scheme was sufficient since absolute phase referencing to the flat mirror is unnecessary to demonstrate the reported phenomena. Some of the measured spectra are shown in Fig. 3 to provide perspective on the general effects of each sample. The plots reveal that the paint, by itself, provides some broadband enhancement of the reflectivity over bare drywall, while the paint doped with fine Al powder provides substantial enhancement of reflectivity. Surprisingly, the paint doped with coarse Al powder only enhanced low frequency reflectivity, and actually reduced reflectivity above about 0.6 THz. Coupled with the lack of noticeable spatial dispersion, this suggests the coarse powder is producing an absorptive effect at higher frequencies, however this is a topic for later studies.

The measurements of the sample with coarse Al powder at the 500 mg/mL (highest) concentration provided the most noticeable and informative results. Four specific measurements that are discussed in the following sections are illustrated in greater detail in Fig. 4. For these measurements, the entire drywall sample was divided into four areas designated C1 through C4, and Fig. 4 shows a white circle in each area to indicate the measured terahertz beam position and size. Each circled spot was measured three times using THz-TDS to ensure repeatability. Since all four spots were measured from the same overall drywall sample, these provided a good demonstration of the variety of measured waveforms that can result from a single texture application.

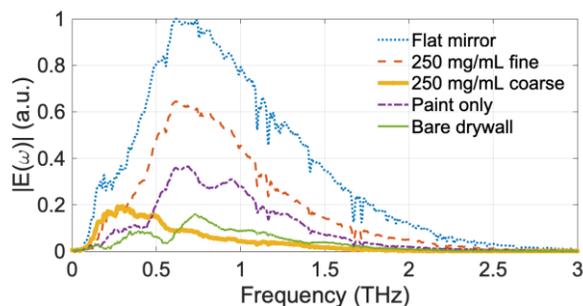

**Fig. 3**: Normalized frequency spectra of various measured terahertz signals. Small downward spikes, most evident on the flat mirror spectrum, were caused by residual water vapor, which was not totally eliminated by the dry-air enclosure around the THz-TDS system.

### III. MEASUREMENTS, MODELING, AND SIMULATION

#### A. Sample Measurements

Spectral reflection data from areas C1-C4 are shown in Fig. 5, obtained by applying the discrete Fourier transform to the measured time-domain terahertz amplitude data. These data warrant some discussion to put them in context of other work.

First, the bandwidth changed considerably among the measurements. Measurement C1 (Fig. 5) had observable components only below 1.5 THz, whereas measurement C2 had observable components to almost 2.5 THz. In addition, the main body of the spectrum (evaluated at full-width-half-max) ranges from about 0.5 THz (C1) to 1 THz (C2). This is quite different from the results shown in Dikmelik *et al.* [20], in which a clear Gaussian frequency roll-off was observed at a particular frequency corresponding to the standard deviation of surface height. This may suggest that our surfaces do not have a purely Gaussian surface height distribution or that they are otherwise more variable than those observed in [20]. Future visible or mechanical quantitative surface topology studies should reveal some of these details.

Second, interference nulls are visible at various locations in the spectral data. These nulls may be the result of frequency selective fading [22] and indicate the ever-present, but random, possibility of poorer signal transmission in certain bands for static channels. Similar to spatial interference patterns observed in earlier work [11], this fading appears to be caused by multipath effects where the roughened wall produces a delay spread within the area of the reflected beam. While some of these nulls are substantial, most are not deep enough to suggest a total loss of signal power to the extent that communications would be impossible. However, due to the analyticity of the signals, each null must be accompanied by GDD. Since these nulls are approximately 100 GHz wide, and can appear almost anywhere in the usable spectrum, it means GDD is always a potential problem for high bandwidth terahertz signals. Likewise, it suggests that the presence of GDD in a particular band of interest is likely to be unpredictable or even stochastic in some cases.

Third, all of the measurements had a low cutoff frequency at approximately 60 GHz. This is not due to the samples but is an inherent characteristic of this particular THz-TDS system. Since all of the measured spectral features relate to GDD and resulting communication system performance, recreating them in simulation was one of the main goals of the model.

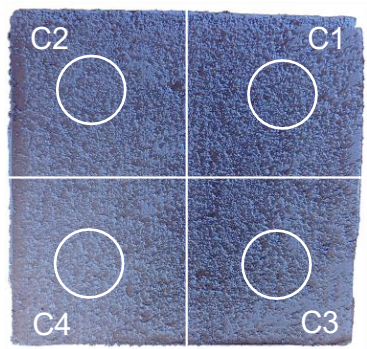

**Fig. 4:** Division of sample shown in Fig. 1, showing the four specific areas illuminated by the terahertz beam (white circles), C1-C4. Each illumination spot was approximately 20 mm in diameter.

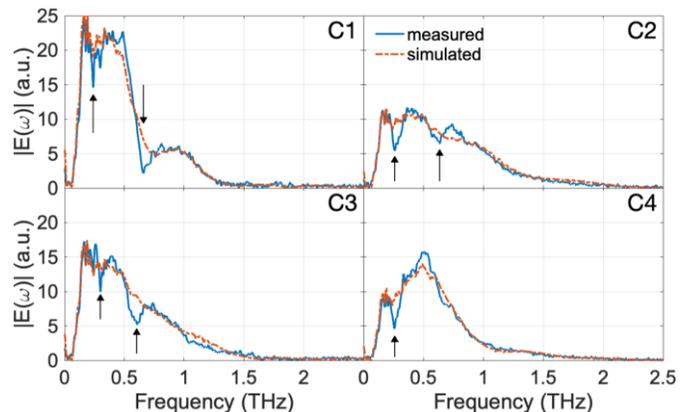

**Fig. 5:** Measured and simulated data from the four beam locations (C1-C4) of the 500mg/mL coarse Al powder sample. Vertical arrows designate interference nulls.

## B. Modeling

Similar to the work of Dikmelik, *et al*. [20] the scattering surface was modeled as a set of planar perfect reflectors, where each element has a random height distribution (in the $\pm y$ direction) relative to the average surface position ($y = 0$). Areas with a greater variance in the size of texturing particles would, in principle, be correlated with models possessing a greater variance in the random distribution of reflector heights. The two-dimensional distribution of these reflectors in the *xz*-plane was ignored, effectively giving each reflector of the set the same position ($x = z = 0$). While this prevents the determination of effects due to *transverse* surface correlation lengths [11] it also isolates observed effects to the time-domain, simplifying the temporal dispersion analysis and mechanisms of causation. It was assumed that the surface height distribution of the textured samples was Gaussian based on previous work [20]. When a terahertz wave reflects from the surface model, the distribution of reflector heights is manifest as a distribution of random time delays, resulting in temporal dispersion (delay spread) in the final waveform.

To simulate the reflection off a rough wall, first an input waveform, $a_o(t)$, is established. We used an actual THz-TDS specular reflection measurement from the flat reference mirror to generate $a_o(t)$. The model then calculates a time delay $t_i$ in the reflected wave for every reflector in the modeled surface height distribution, and these times shifts are then applied to copies of the input pulse. The resulting delayed reflected waves are finally averaged together to make the final output waveform, $a(t)$ where $N$ is the number of reflectors in the set,

$$a(t) = \frac{S}{N}\sum_{i=1}^{N} a_o(t - t_i). \quad (1)$$

A global scaling factor $S$ accounts for losses due to diffuse scattering, material absorption, and general system inefficiencies, however it also presumes that each reflected wave of the set has the same signal magnitude.

## C. Simulation

The model was tested by simulations using a set of $N = 100$ reflectors with random time-delays. One thousand simulations were performed, each with a different random distribution of time shifts $t_i$ across the set, to determine the variability in observable reflection spectra. The time shifts were distributed around some average $t_0$ according to Gaussian statistics with a specified standard deviation according to the random height of the reflectors. To illustrate, four example frequency spectra obtained from the 1,000 simulations are shown in Fig. 5. As in the measured data, these spectra were obtained by Fourier transformation of the simulation's time-domain output. The model is capable of producing a much wider variety of responses; however, these were selected due to their similarity to actual measured data from four different areas of the textured drywall.

While an exact fit to the data is not expected from this approximate stochastic model, many of the experimental features appear to be easily reproducible, including the spectral magnitudes, bandwidths, general spectrum shape, and some interference nulls. As such, we regard this scattering (channel) model adequate for follow-on studies of GDD and its effect on communication system performance. We note that, apart from adjusting the scaling factor $S$ to normalize out system losses, no attempt was made to fit the model spectra to the measured data. The matches shown are purely the result of different random distributions of reflectors and their associated temporal dispersion, suggesting that the original assumption of a Gaussian height distribution is reasonably accurate. The simulations shown were performed using the Gaussian distribution mean and standard deviation parameters listed in Table 1.

TABLE 1

| Sample | C1 | C2 | C3 | C4 |
|---|---|---|---|---|
| Mean (µ) | 150 µm | 225 µm | 175 µm | 275 µm |
| Standard Dev. (σ) | 140 µm | 160 µm | 175 µm | 125 µm |

## IV. GROUP DELAY DISPERSION AND SYMBOL ERROR RATES

### A. Group Delay / Dispersion

The effects of multipath scattering are well known in wireless systems but have been studied relatively little in the terahertz regime thus far. Fading (small-scale) is often described as the principal penalty from multipath propagation. By the analyticity of the waves, any frequency-dependent amplitude effect in fading must be accompanied by some frequency-dependent change in phase (time shift and temporal dispersion). This effect on phase becomes increasingly impactful with higher bandwidth signals such as would be expected in next-generation terahertz wireless systems. Our model can be used to quantify these effects by employing the concepts of group delay (GD) and GDD, which are the first and second derivatives, respectively, of the wave phase $\varphi$ with respect to angular frequency.

$$GD = \frac{\partial \varphi}{\partial \omega} \qquad GDD = \frac{\partial^2 \varphi}{\partial \omega^2} \quad (2)$$

When the GD is variable over frequency GDD becomes non-zero, or equivalently there is non-zero second-order dispersion in the channel that spreads out temporal waveform features [23]. Fig. 6 shows four examples of the GDD predicted by our model when terahertz waves reflect off random textured surfaces with standard deviation in height of σ =300 µm. Since the model essentially disregards transverse particle extent, the mean value of the particle size has no dispersive effect and only produces a constant phase shift, which contributes nothing to GD or GDD. The plots confirm that GDD can reach substantial levels over bandwidths approaching 100 GHz. Again, given the random nature of texturing, this large band of dispersion can occur almost anywhere between 0.3-1.0 THz. Coarser textures (larger σ) exhibit stronger GDD bands at lower frequencies. Substantial GDD effects begin to occur once σ of the surface becomes comparable to the operating wavelength, since such a surface would impart substantial and variable phase shifts to the various frequency components in the spectrum.

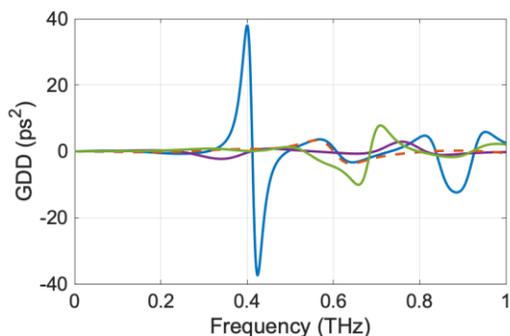

**Fig. 6:** Four GDD plots resulting from different random instances of the same distribution. These are simply random GDD examples produced by the model and are not correlated with the measurements in Fig. 5.

Although plots such as Fig. 6 are useful in demonstrating the varied types of GDD profiles that can arise from a textured surface, they are not well suited to predicting the precise impact of such a surface on the fidelity or data-carrying capacity of an incident communication signal. To determine whether the GDD presented in Fig. 6 is truly "substantial," it is necessary to quantify the expected SER – that is, the probability that reflection from the surface will cause numerous transmitted communication symbols to be received in error. For a given data rate, temporally dispersing the waveform via GDD causes neighboring bits in a digital data stream to blend into one another, resulting in inter-symbol interference (ISI) [24]. Therefore, as GDD increases, ISI also increases, resulting in a larger SER. In this way, GDD directly impacts what SERs are achievable for a particular data rate. Conversely, GDD will impart a reduced upper limit on data rates for a fixed low SER, even in the absence of noise [25], which is detrimental since the primary motivation of terahertz wireless is to achieve extremely high data rates.

To quantify how scattering from rough surfaces affects the relationship between data rate and SER, a communication system was simulated at 0.25 THz. The simulation generates a pseudo-random symbol stream, splits the data into Q and I channels, and then modulates the two channels onto the carrier at 0.25 THz, using phase shift keying. This results in a quadrature phase shift keying (QPSK) modulation scheme, which offers 2 data bits per symbol. The signal is Fourier transformed and the resulting spectrum is multiplied by the channel (scattering surface) transfer function, which was generated by our model with a random surface height distribution, as explained in Section 3. The output spectrum is then converted back to a time-domain signal for sampling and demodulation. The received data is compared to the transmitted data to calculate the SER. Symbol rates of 10, 30 and 50 GBd were chosen to test the channel. Additive white Gaussian noise was added at the receiver to produce a signal-to-noise-ratio (SNR) of 50 dB as measured at the receiver, implying the SNR is not a limiting factor. The simulation runs iteratively over specific standard deviations one at a time. At each standard deviation, a new random reflection surface is generated by our scattering model. For each surface, the simulation then sends approximately 4.4 million pseudo-random symbols through the channel and calculates the SER. This entire process was repeated 122 times, each time with new randomized surfaces.

Finally, all the calculated SER results were averaged to produce the presented SER results.

Figure 7 shows the averaged SERs and how they increase as the standard deviation of scatterer size (surface roughness) increases. It is of particular importance that symbol errors are almost non-existent up to a certain surface roughness (deemed the "dispersion limit"), at which point they abruptly turn on and then quickly reach levels that would render the communication system almost unusable. This threshold appears to occur when $\sigma \sim 0.30\lambda$ -$0.33\lambda$. As expected, higher symbol rates result in worse SER, since bits would be spaced more closely together in time, giving greater opportunity for ISI. While this doesn't appear to change the dispersion limit significantly, it does result in orders of magnitude higher SER. The large SNR of 50 dB means these effects are almost entirely due to dispersion alone. Lower SNRs would be expected to produce even greater SERs. Upon preliminary investigation, the jagged appearance of the simulation data appears to be due to the random occurrences of large GDD spikes (like the blue curve in Fig. 6) coinciding with the simulated channel bandwidth. Such spikes produce a large number of errors, which are not easily averaged out by subsequent simulations and therefore remain after the entire simulation process is complete. In addition, they only ever make SER worse, hence the upward pointing nature of the spikes on the SER plots. As is apparent in Fig. 7, the lower data rate links (e.g. 10 GBd) are more severely affected by this phenomenon because their operational bandwidth is narrower, which means the entire bandwidth of the link may be strongly affected by GDD.

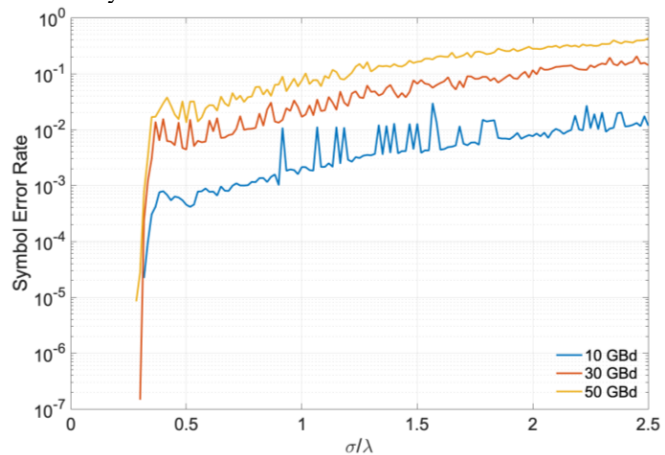

**Fig. 7:** Symbol error rates produced by single reflections from a random diffuse surface whose height distribution standard deviation σ is expressed in fractions of a wavelength λ on the abscissa. Data is QPSK modulated at a 0.25 THz carrier frequency with a received SNR of 50 dB. Results shown are an average of 122 simulations from different random surface instances.

Figure 7 showed the effect caused by the dispersion resulting from a single reflection from a diffuse surface. Such effects, however, will accumulate when multiple reflections occur, which is important in the design of NLOS links. Figure 8 shows the effects to a 50 GBd, QPSK modulated data stream using a 0.25 THz carrier with an SNR of 50. This simulation was performed in the same manner as the previous simulation with the following exceptions: the simulation was averaged 200 times and employed only about 1.1 million pseudo-random symbols per simulation. For each reflection, a new random surface was generated. The effect of all the reflections is finally determined by the product of all the associated channel transfer functions. Multiple reflections are observed to shift the dispersion limit to smaller σ, meaning the channel is becoming more sensitive to the texture of the surfaces. At 5 reflections, which is here assumed to be the maximum encountered in practical NLOS systems, the dispersion limit has reduced to σ ~ 0.23λ-0.25λ. Moreover, the SER beyond that limit is truly abysmal, quickly reaching 0.50 and trending to 0.75 (3 of every 4 symbols is in error) as σ grows to multiple wavelengths. An SER = 0.75 corresponds to what would be expected by comparing completely distinct random symbols with the original transmitted symbol stream, implying no information is being conveyed at all.

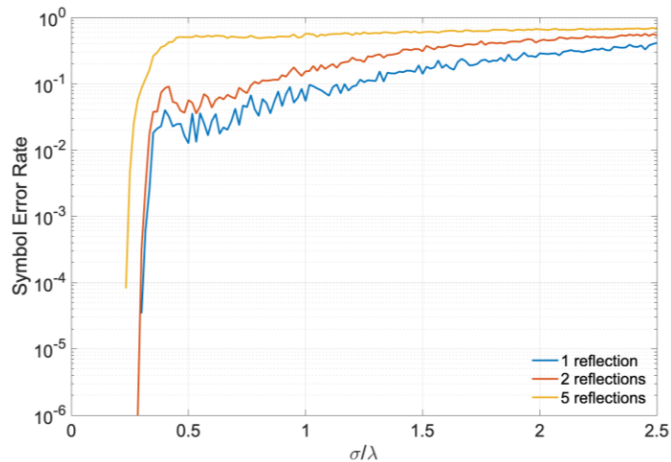

**Figure 8:** Symbol error rates produced by multiple reflections from random diffuse surfaces whose height distribution standard deviation σ is expressed in fractions of a wavelength λ on the abscissa. Data is QPSK modulated at a 0.25 THz carrier frequency with a received SNR of 50 dB. Results shown are an average of 200 simulations from multiple random surface incidences.

## V. Conclusion

This paper shows that the texture or roughness of reflecting surfaces could become a major limiting factor in the error rate performance of NLOS, high data rate terahertz wireless communication systems. Though diffuse surfaces are promising for spatial signal dispersion, facilitating signal coverage, our work indicates that these surfaces also cause temporal dispersion, specifically GDD. Moreover, this GDD results in an unpredictable form of ISI that abruptly turns on symbol errors, increasing their rate by orders of magnitude as the dispersion limit of surface roughness is passed. This phenomenon is not due to reduced SNR and therefore could not be compensated by increasing signal power. For high bandwidth applications, the increasing error rate caused by ISI, rather than signal losses, could become the limiting factor to achievable data rate. This work also shows that each reflection from a rough surface will further increase the error rate of the system, meaning multi-reflection NLOS links are affected even more greatly. The presented results suggest that future terahertz wireless systems may need to compensate for this dispersion, or that high bandwidth terahertz applications will be limited to LOS in some scenarios.

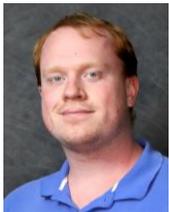
**Russ Messenger** received his B.Sc. degree in electrical engineering from Oklahoma State University (OSU), Stillwater OK, in 2019. Currently, he is currently pursuing his M.S. degree in electrical engineering under his advisor John F. O'Hara. He currently is working as a graduate research assistant at the Ultrafast Terahertz and Optoelectronic Laboratory, School of Electrical and Computer Engineering at OSU. His current research interests are in Internet of Things, terahertz band wireless communication, and terahertz time domain spectroscopy.

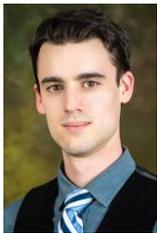
**Karl L. Strecker** received the B.S. and M.S. degrees in electrical engineering from Oklahoma State University, Stillwater, Oklahoma, in 2018 and in 2020 respectively. He is currently pursuing the Ph.D. degree in electrical engineering at the same institution. From 2018 to 2020, he was a Research Assistant in the Ultrafast Terahertz and Optoelectronics Laboratory at Oklahoma State University. His research interest includes wireless terahertz communication systems, group velocity dispersion management, and terahertz material characterization. Mr. Strecker is a recipient of the 2020 National Science Foundation Graduate Research Fellowship.

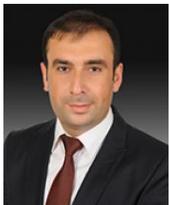
**Sabit Ekin** (M'12) received the B.Sc. degree in electrical and electronics engineering from Eskişehir Osmangazi University, Turkey, in 2006, the M.Sc. degree in electrical engineering from New Mexico Tech, Socorro, NM, USA, in 2008, and the Ph.D. degree in electrical and computer engineering from Texas A&M University, College Station, TX, USA, in 2012. In summer 2012, he was with the Femtocell Interference Management Team in the Corporate Research and Development, New Jersey Research Center, Qualcomm Inc. He joined the School of Electrical and Computer Engineering, Oklahoma State University, Stillwater, OK, USA, as an Assistant Professor, in 2016. He has four years of industrial experience from Qualcomm Inc., as a Senior Modem Systems Engineer with the Department of Qualcomm Mobile Computing. At Qualcomm Inc., he has received numerous Qualstar awards for his achievements/contributions on cellular modem receiver design. His research interests include the design and performance analysis of wireless systems in both theoretical and practical point of views, visible light sensing, communications and applications, non-contact health monitoring, and Internet of Things.

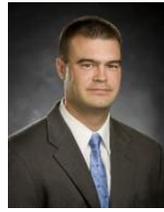
**John F. O'Hara** (M'05 - SM'19) received his BSEE degree from the University of Michigan in 1998 and his Ph.D. (electrical engineering) from Oklahoma State University (OSU) in 2003. He was a Director of Central Intelligence Postdoctoral Fellow at Los Alamos National Laboratory (LANL) from 2004 until 2006. From 2006-2011 he was a Technical Staff Member with the Center for Integrated Nanotechnologies (LANL) and worked on numerous metamaterial projects involving dynamic control over chirality, resonance frequency, polarization, and modulation of terahertz waves. In 2011, he founded a consulting/research company, Wavetech, LLC specializing in automation and IoT devices. In 2017 he joined OSU as an assistant professor in the School of Electrical & Computer Engineering. His current research involves terahertz wireless communications, terahertz sensing and imaging with metamaterials, IoT, and light-based sensing and communications. He has 3 patents and around 100 publications in journals and conference proceedings. Dr. O'Hara is a Senior Member of the IEEE.